\documentstyle[11pt,newpasp,twoside,epsf]{article}
\def\edcomment#1{\iffalse\marginpar{\raggedright\sl#1\/}\else\relax\fi}
\reversemarginpar

\begin{document}
\title{High Dynamic Range and the Search for Planets}
 \author{A. T. Tokunaga, C. Ftaclas, J. R. Kuhn, and P. Baudoz}

\affil{Institute for Astronomy, Univ. of Hawaii, 2680 Woodlawn Dr., Honolulu, HI  96822}

\begin{abstract}
General arguments for optimized coronagraphy in the search for planets
are presented.  First, off-axis telescopes provide the best telescopic
platforms for use with coronagraphy, and telescope fabrication
technology now allows the fabrication of such telescopes with diameters
of up to 6.5 m.  We show that in certain circumstances a smaller
telescope with an off-axis primary has a signal-to-noise advantage
compared with larger Cassegrain telescopes.  Second, to fully exploit
the advantages of the coronagraph for suppressing stray light, it is
necessary to use a high Strehl ratio adaptive optics system.  This can
be best achieved initially with modest aperture telescopes of 3--4 m
in diameter.  Third, application of simultaneous differential imaging
and simultaneous polarimetric techniques are required to reach the
photon-limit of coronagraphic imaging.  These three developments, if
pursued together, will yield significant improvements in the search
for planets.

\end{abstract}

\section{Introduction}

There has been much interest recently in the pursuit of planet
detection using 8 and 10 m telescopes, primarily to take advantage of
the higher angular resolution and greater light gathering power.  The
primary problem to overcome is the large amount of scattered light
from the star in a region where planets might be found (within
1--2\arcsec\ radius of the star).  Since the reflected light from a
Jovian-size planet is about 10$^{-9}$ that of the primary star's
intensity, it is necessary to employ techniques to reduce the
scattered light arising from Earth's atmosphere, secondary mirror
spiders, the hole in the primary mirror, and from optical surfaces
such as the primary and secondary mirrors of the telescope.\\

In this paper we review the advantages of off-axis telescopes and the
advantages of searching for planets with a high-order adaptive optics
(AO) systems on 3--4 m class telescopes.  It is worth noting that some of the
most significant astronomical discoveries in the past decade were made
using modest aperture telescopes (about 2 m or smaller), including

\begin{itemize}
	\item The discovery of the first brown dwarfs (Nakajima et al. 1995; 
Rebolo et al. 1995)
	and the subsequent discovery of field brown dwarfs by the 2MASS and 
	DENIS surveys.
	\item The discovery of planetary mass objects (Mayor \& Queloz 1995)
	\item The discovery of Kuiper Belt objects (Jewitt \& Luu 1993)
\end{itemize}

We suggest here an approach for the direct detection of
planets by using modest aperture telescopes to develop
techniques that can be scaled up to larger telescopes.  \\

\section{Advantages of Off-Axis Telescopes}

A case for off-axis telescopes for research requiring high dynamic
range has been made by Kuhn \& Hawley (1999).  Here we summarize a few
major points.  Figure 1 shows the scattered light from a Cassegrain
telescope.  It illustrates the problems in attempting to detect
planets near a bright star, namely saturation of the CCD (the vertical
stripe), scattered light from the secondary spiders (cross feature at
45\deg), and scattered light from the atmosphere and mirrors.  \\

\begin{figure}[hb]
\plotone{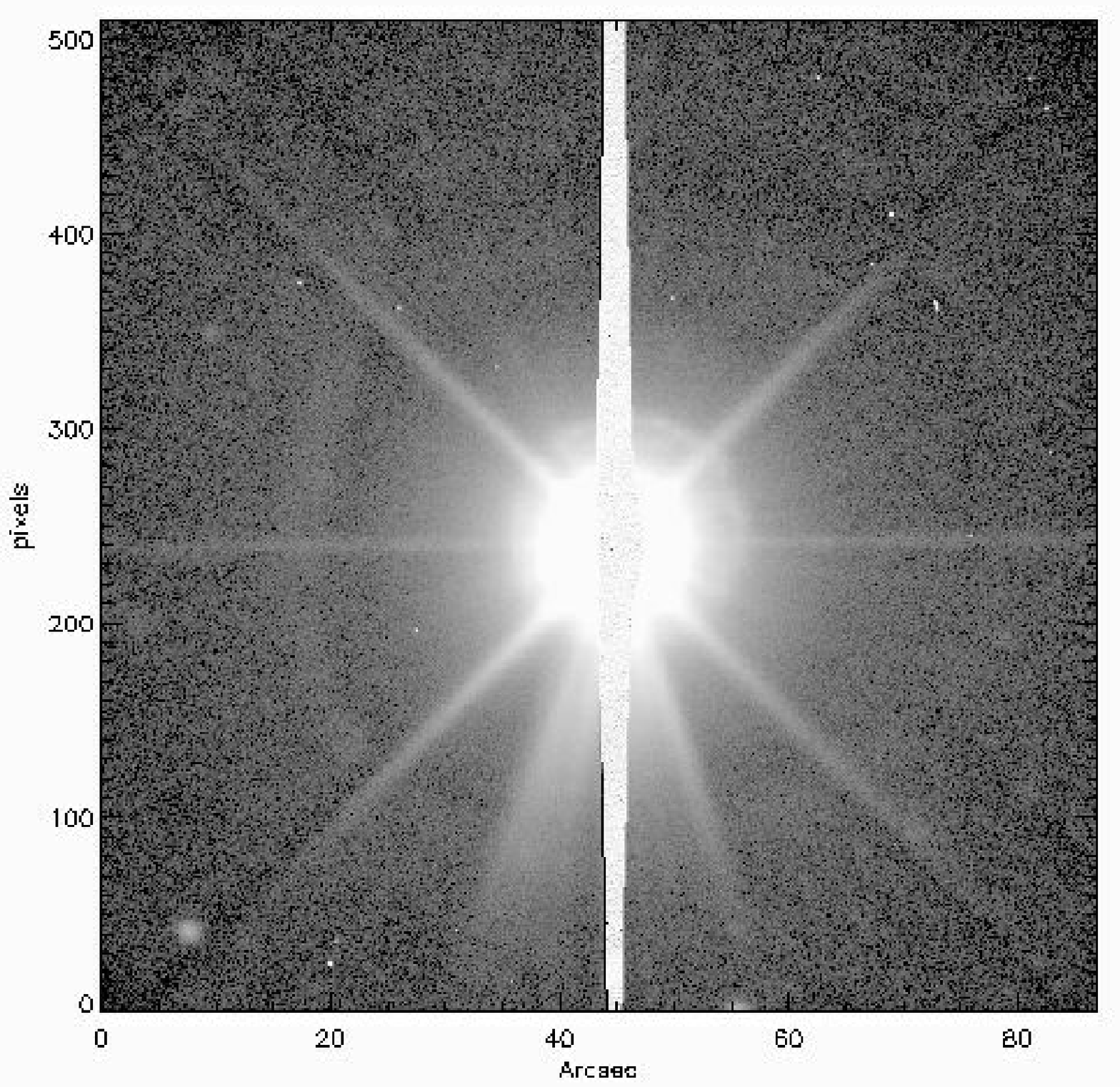}    
\caption{Scattered light from the Blanco telescope.  Adapted from 
Kuhn \& Hawley (1999).}
\end{figure}

Use of an occulting disk in the focal plane will eliminate the
saturation effects of the point source.  For the sources of scattered
light Kuhn \& Hawley (1999) argue that an off-axis telescope will
provide a greatly improved point-spread function (PSF).  This is
illustrated in Figure 2.  The off-axis telescope provides an
unobscured pupil image, thus eliminating scattered light components
from the secondary spiders and the central hole in the primary mirror. 
In the search for planets, the relatively clean PSF of the off-axis
telescope is a great practical advantage.  \\

\begin{figure}[hb]
\plotone{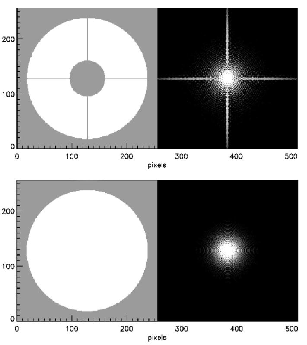}    
\caption{Comparison of obscured and unobscured pupil images.
Adapted from Kuhn \& Hawley (1999).}
\end{figure}

In attempting the direct imaging of planets, it is desirable to employ
a coronagraph to suppress the light from the star.  A comparison of
the coronagraphic performance of a Cassegrain and off-axis telescope
is shown in Figure 3.  The calculation assumes scattering from the
atmosphere and natural seeing (no AO).  Although the gain is modest
(about 0.5 mag at the most), it should be noted that the scattered
light from the secondary spiders is significant (see Figs.  1 and 2). 
We also show in Section 3 that further sensitivity gains can be made
by using a high-order AO system.  \\

\begin{figure}[t]
\plotone{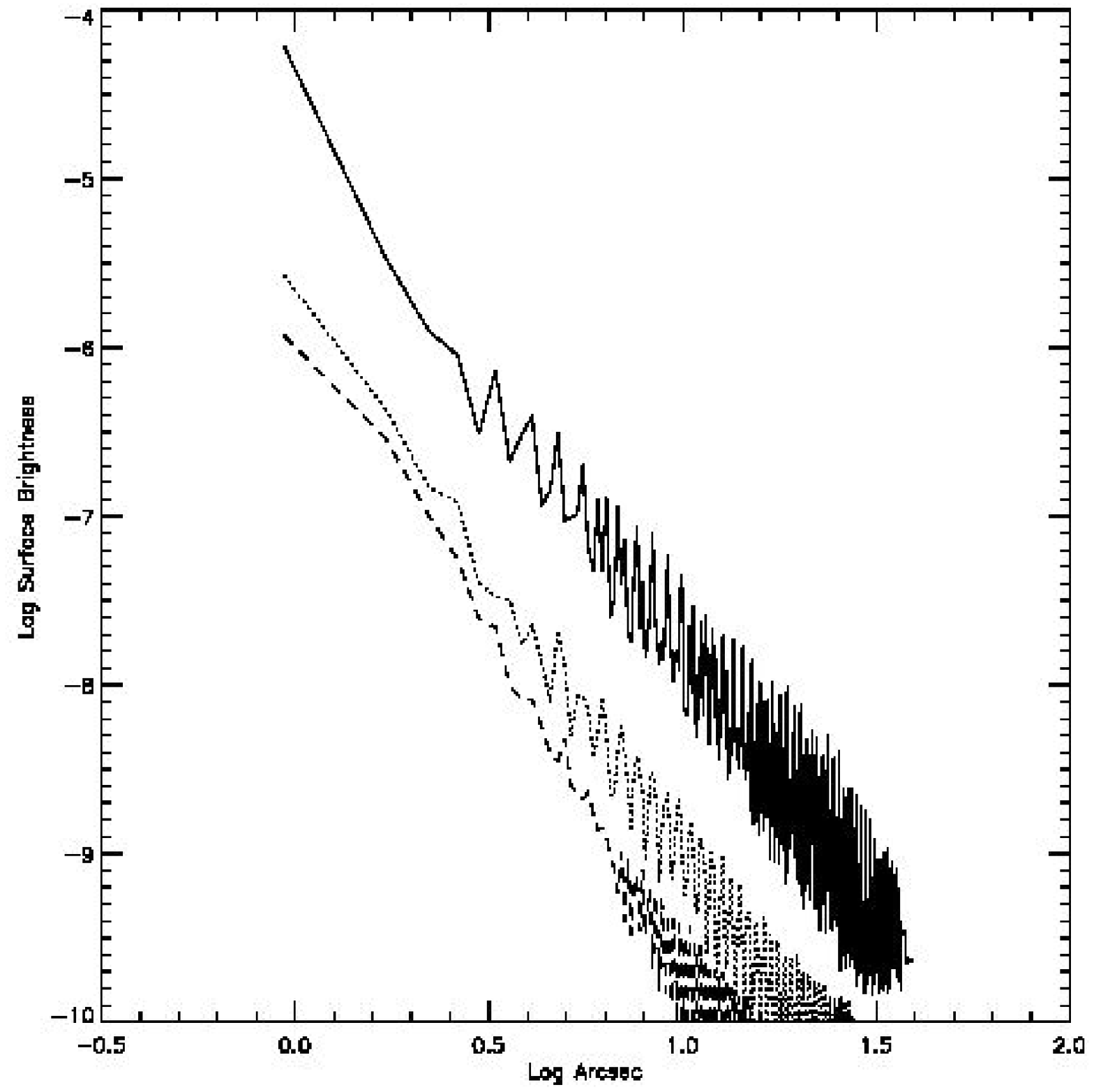}    
\caption{Comparison of coronagraphic performance of a Cassegrain 
telescope to that of an off-axis telescope.  The upper curve shows the 
radial surface brightness profile of the direct image, the middle 
curve shows that of a Cassegrain coronagraphic PSF, and the lower curve 
shows that of an off-axis telescope coronagraphic PSF. 
Adapted from Fig. 14 of Kuhn \& Hawley (1999).}
\end{figure}

There is a further advantage of the off-axis telescope with a
coronagraph that arises from the significantly higher throughput in
the off-axis telescope case.  There are two items that affect the
throughput.  The first is that the encircled energy of the central
core is less for a telescope with a central obscuration than one
without a central obscuration.  The second is that the throughput of a
coronagraph for a cassegrain telescope is less than that of an
off-axis telescope.  These effects combine to make the integration
time required to detect a faint nearby planet with an off-axis 4 m
telescope competitive with that of an 8 m Cassegrain telescope for
coronagraphy without adaptive optics.  This is demonstrated below.  \\

A Cassegrain telescope has an encircled energy in the
diffraction-limited image core that decreases with the size of the
linear obscuration of the secondary (Schroeder 1987).  In Table 1 we
show the encircled energy in the image core for an off-axis 4-m
telescope, an off-axis 6-m telescope, and a Cassegrain 8-m telescope
with a secondary having a linear obscuration of 0.25 (a 2 m diameter
secondary).  We show the encircled energy before the pupil mask of the
coronagraph and after the pupil mask.  A pupil mask that is 0.9 times
the diameter of the primary mirror is assumed.  Considering that the
hole in the primary mirror must also be masked, the central
obscuration increases from 0.25 to 0.40 in the case of the Cassegrain
telescope, and the encircled energy is correspondingly reduced to
0.55.

\begin{table}
\caption{Comparison of encircled energy \\}
\begin{tabular}{c|c|c|c|c|c}

\tableline
\multicolumn{3}{c}{Before Lyot Mask} \vline & 
  \multicolumn{3}{c}{After Lyot Mask}\\
\tableline
Diameter  & Linear      & Encircled  & Diameter & Linear      & Encircled \\
 (m)      & Obscuration & Energy     &   (m)    & Obscuration & Energy \\
\tableline

4        & 0.0         & 0.84       & 3.6      & 0.0         & 0.84 \\
6        & 0.0         & 0.84       & 5.4      & 0.0         & 0.84 \\
8        & 0.25        & 0.70       & 7.2      & 0.4         & 0.55 \\
\tableline
\tableline

\end{tabular}
\end{table}

The reduction in encircled energy lowers the signal-to-noise in the
detection of planets, that is, the peak flux in the planet PSF is
reduced.  Assuming background-limited imaging with AO, the integration
time is proportional to $~\sim 1/(A \phi)^{2}$, where $A$ is the
effective collecting area and $\phi$ is the peak object
intensity/background intensity.  Table 2 shows the integration time
ratios for a 4 m off-axis telescope to an 8 m Cassegrain telescope and
for a 6 m off-axis telescope to an 8 m Cassegrain telescope.  We see
that after the coronagraph, the 4 m telescope requires 4 times longer
integration, but this is much less time than the factor of 11 times
longer before the coronagraph.  The 6 m off-axis telescope is better
than the 8 m cassegrain telescope after the coronagraph.  \\

In the case of background-limited imaging and no AO, the integration
time is proportional to $~\sim$$1/(A \phi^{2})$, and the ratio of
integration times is shown in the column 4 of Table 2.  In this case
the 4 m off-axis telescope is competitive with an 8 m Cassegrain
telescope (except for the smaller diffraction-limited core of the 8 m
telescope).  These calculations also do not consider the effect of the
scattered light from the secondary spiders, as mentioned above.  These
arguments were previously made by Ftaclas (1994).  We therefore
conclude that using off-axis telescopes for the detection of planets
should be seriously considered.

\begin{table}
\caption{Comparison of integration time \\}
\begin{tabular}{c|c|c|c}

\tableline
Integration & Before      & After       & No Adaptive \\
Time Ratio  & Coronagraph & Coronagraph & Optics      \\
\tableline
t(4m)/t(8m) & 11          & 4           & 1.3   \\
t(6m)/t(8m) & 2.2         & 0.74        & 0.75  \\
\tableline
\tableline

\end{tabular}
\end{table}

\section{The Case for High-Order AO on ``Small'' Telescopes}

Lai (2001) and Oppenheimer, Sivaramakrishnan, \& Makidon (2002) make
the case for pursuing high-order AO on 3--4 m class telescopes.  We
follow the arguments of Oppenheimer et al.  here, since they address
specifically the question of detecting planets.  For a given number of
actuators on the AO system, higher spatial frequencies will be
corrected by the AO system on a smaller telescope than on a larger
telescope.  Therefore the Strehl ratio (SR) will be higher on the
smaller telescope.  The contrast improvement is
$$
\eta = \frac{ S_{\rm{AO}} }{ S_{\rm{see}} } (1 - S_{\rm{AO}})^{-1} ,
$$

\noindent where S$_{\rm{AO}}$ is the SR of the AO system and
S$_{\rm{see}}$ is the SR of the seeing limited image.  The first term
is the increase in the contrast of the image, and the second term is
the increase in the energy from the wings to the core.  It is evident
that the contrast improvement is an extremely sensitive function of
the SR of the AO system.  Oppenheimer et al.  argue that a very high
SR on a 3--4 m telescope opens up an observational niche between the
direct imaging of planets without adaptive optics and the indirect
imaging of planets using Doppler-shift methods.

What SR is required to achieve this observational niche?  Oppenheimer
et al.  find that for a 3.6 m telescope, a SR of 95\% is a desirable
goal.  To reach this high SR,  approximately 34
actuators across the aperture of the telescope (assuming a
Shack-Hartmann wavefront sensor) are required.  More actuators would
begin to decrease the SR because there would be less photons for each
element of the wavefront sensor.  Furthermore, other effects such as
atmospheric scintillation and speckle noise can become serious
limitations at very high SRs.  The former has not been addressed in a
practical system as yet, whereas the latter is presently being
addressed in several astronomical systems, as discussed in 
Section 4.

There is another important advantage of a high-order AO system.  As
the number of actuators increases, the AO system will be able to
correct higher frequency wavefront errors.  The spatial frequency
cutoff is given by $k_{\rm{AO}} = N_{\rm{act}}/2D$, where
$N_{\rm{act}}$ is the number of actuators across the diameter of the
primary mirror and $D$ is the diameter of
the telescope.  In the image plane this corresponds to an angle on the
sky, $\theta_{\rm{AO}} = N_{\rm{act}}\lambda/2D$, over which the AO
system is able to provide wavefront correction.  This is illustrated
in Figure 4.  The most critical region for searching for planets
around nearby stars would be at angular distances less than 2\arcsec\
from the star, a region that is mostly covered with $N_{\rm{act}} = 34$
on a 3.6 m telescope.

\begin{figure}
\plotone{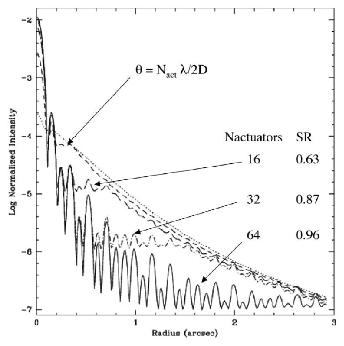}    
\caption{Radial profiles of simulated AO PSF (assuming a 
Shack-Hartmann wavefront sensor).  $N_{\rm{act}}$ is the number of
actuators across the primary mirror.  Adapted from Oppenheimer et al. (2002).
See also Sivaramakrishnan et al. (2001).}
\end{figure}

Note that Lai (2001) presents simulations that show curvature sensing AO 
systems can achieve a SR $> $90\% at K with as little as 104 actuators on a 
3.6 m telescope.  This suggests that further development of curvature
systems is warranted.

Since a coronagraph typically has an occulting disk that is around 
$4\lambda/D$ in diameter or larger, the region that is accessible with 
AO wavefront correction can be very small if $N_{\rm{act}}$ is small or if 
the telescope aperture is large.  Present coronagraphic systems have 
not met expectations through a combination of $N_{\rm{act}}$ being too 
small or not having a way to reduce speckle noise.  We can address the 
former problem by building a high-order AO system on a modest aperture 
telescope, but we must still deal with the speckle noise problem which is
discussed in Section 4.

\section{Further Scattered Light Suppression}

One must achieve photon noise limited imaging in the search for
planets.  In practice, AO observations are limited by speckle noise,
the noise resulting from the residual uncorrected wavefront errors. 
Techniques have therefore been pursued to employ differential imaging
to reduce the speckle pattern by subtracting images obtained
simultaneously at two wavelengths.  This technique relies on the
search for spectral features that are different at the two wavelengths
so that a residual signal from the planet can be detected (see Ftaclas
2003).  A further refinement of this technique using three wavelengths
is discussed by Marois, Doyon, \& Nadeau (2003).  Practical problems
include small differences in the speckle pattern resulting from
differences in the wavelengths, so it is still not possible to reach
the photon noise limit.

Another approach is to seek a polarization signature by observing all 
of the Stokes parameters simultaneously, as demonstrated by Kuhn, Potter, 
\& Parise (2001a).  They show the feasibility of this approach.  More 
extensive observations using a 36-element AO system on the Gemini 
North Telescope has shown that one can approach the photon-noise 
limit (Potter, private comm.).

One or both of these techniques employed with a coronagraph will have
to be employed to detect planets.  It is critical to reach the 
photon-noise limit, and further work to demonstrate that this
can be achieved is required.

\section{A Development Path}

What is required to take the next step is to combine these ideas into
a practical system.  There is already a N$_{\rm{act}} = 34$ system on
the 3.7 m AEOS telescope on Haleakala.  This could be exploited if the
performance of the system is as good as predicted and a coronagraph is
built to exploit it (Oppenheimer et al.  2003).  Alternatively a 3--4
m class off-axis telescope with a high order AO system could be built
to pursue the ideas presented here.  The practical limitations of
speckle noise and atmospheric scintillation would have to be overcome,
but such a facility could demonstrate techniques to achieve photon
noise limited imaging with a coronagraph and take advantage of the
observational niche for searching for planets as described by
Oppenheimer et al.  (2002).

It is clear that going to larger off-axis telescopes would have the
advantages of the larger collecting area and smaller
diffraction-limited core.  The calculations shown in Table 2 show that
a 6 m off-axis telescope would be superior to a Cassegrain 8 m
telescope on integration time considerations alone.  It is also clear
that an off-axis telescope would have the advantage of a cleaner PSF,
as demonstrated in Figure 2.  Therefore the next logical step would be
to go to the 6 m size to fully exploit the advantages described here
with a single aperture telescope.  Such a facility would be the
premier ground-based infrared telescope because an unobscured aperture
would also provide minimum telescope emissivity.

In the future, it may be possible to build a multiple aperture large
telescope using off-axis mirrors which would have advantages over
large segmented mirror telescopes.  This is discussed in detail by
Kuhn et al.  (2001b).

\section{Summary}

We have summarized ideas that could significantly advance the
detection of planets around nearby stars.  Realization of these ideas
would allow observations near bright stars that are limited by photon
noise rather than scattered light and speckle noise.  There are two
key points:

\begin{enumerate}
\item Unobscured apertures (off-axis telescopes) provide higher 
throughput and a cleaner PSF.

\item High-order AO systems greatly enhance the performance of 
coronagraphs. Implementation of high-order AO systems on modest 
aperture telescopes (3--4 m) will initially provide an observational 
niche for detecting planets.  Techniques to suppress speckle noise 
will be required to approach the photon noise limit.  
\end{enumerate}

\end{document}